# EXPLORING MULTI-DIMENSIONAL EVENTS CHARACTERIZING TECH START-UP EMERGENCE IN THE NIGERIAN ENTREPRENURIAL ECOSYSTEM

Emmanuel Okoro Ajah, American University of Nigeria, Yola, Nigeria, emmanuel.ajah@aun.edu.ng

Chidi Ononiwu, American University of Nigeria, Yola, Nigeria, chidi.ononiwu@aun.edu.ng

**Abstract:** Most countries across the globe identify technology-based start-ups as a driving force for job creation, economic growth and national development, and a critical tool for economic sustenance during pandemic crises like covid-19. However, its emergence are been argued to be problematic. Especially in a developing economy like Nigeria, where tech start-up founders are faced with diverse form of constraints and environmental uncertainties. Extant literature indicated that studies are been conducted to explain tech start-up emergence. However, such studies are fragmented with findings that are determinants to tech start-up emergence, with several determinants studied in isolation, and the emergence as linear and unidimensional events. Consequently, neglecting multi-dimensional perspective, which aggregate the dimensions of events characterizing tech start-up emergence. Given the iterative, event-based process, and interactive-dependent nature of tech start-up ventures to create activity-based products/services in an open, uncertain, nonlinear and dynamic environment, we argue that little are been known about tech start-up emergence. Thus, by drawing from synthesize literature review, activity theory, and exploratory case study design we identify opportunity discovery and selection; team formation and domain consensus; bootstrapping; minimum viable product development and market experimentation feedback as interdependent multi-dimensional events constituting tech start-up emergence in Nigerian tech start-up ecosystem.

## 1. INTRODUCTION

It is evident globally that technology-based start-ups (hereafter-called tech start-ups), is a driving force for job creation and economic growth and as such, considered as socio-economic tool and mechanism for national development. Kirkley (2016) describe it as a tool that indicate the state of economic health and prosperity of a nation. With focus on Nigerian tech start-up ecosystem, a community of interdependent and interactive actors (Motoyama & Knowlton, 2017), with significant attractive market, and ranked one of the top three tech start-up ecosystem in sub-Saharan region (David-West et al., 2018; Onoja, 2020). Recently in Nigeria, tech start-ups played a critical role in contributing to economic sustenance and well-being of the people during the year 2020 Covid-19 pandemic lockdown. During the lock down, most economic activities were shut down for months. However, Nigeria tech start-up ecosystem were busy, as many tech start-ups emerged during this period, with the aim of proffering solution to essential services that were disrupted by lockdown. For instance, uLesson platform was founded in 2019 to help students at home during the pandemic, it provided learning solution that is of high quality, affordable, and easily accessible to students in Nigeria and across Africa (Jackson, 2021). Consequently, tech start-ups as a digital social enterprise became a rescue medium for people in Nigeria to survive the covid-19 pandemic lockdown challenges.

However, literature indicated that failure of tech start-ups abounds, though with few successful ones. Examples of failed tech start-ups in our context include OLX, Efritin, WeChat, Easy Taxi (Nairametrics, 2020). Nigerian OLX tech start-up, a vibrant new venture that matches demand and supply of products was short down in 2018; Secondly, Easy Taxi, a popular taxi hailing tech start-up also closed shop in 2016 (Nairametrics, 2020). Nascent entrepreneurs involved in the activities of developing tech start-ups (Diakanastasi & Karagiannaki, 2016), consistently "contend with extreme uncertainty in the early phases of development. This "creates unforeseen contingencies as conditions evolve" (Townsend, 2012, p. 151 ). The alarming failures, suggest that tech start-ups gestation process is problematic (Reymen et al., 2015). Gestation process is a development process that spans through ideation to commercialization in start-up life cycle (Reynolds & Miller, 1992). This argument suggests that start-ups struggles for existence against prevailing uncertainties, as they go through gestation (Salamzadeh & Kesim, 2015). Such uncertainties manifest in the form of poor infrastructures, unfavorable government policies, duplicate government taxes, even in the absence of amenities that the payment of such taxes should have provided due to corruption, political instability, and lack of investors (Erik, 2013).

Thus, the unequal flux of low success to high failure rate experienced in tech start-ups has attracted the attention of scholars, professionals, policy makers, across the globe. Consequently, its negative effect on economic growth and national development is alarming, thus, we argue that it is critical to understand necessary actions required to develop a viable tech start-up (Park, 2005). Corroborating this argument, Yang et al. (2017) suggest the need to further investigate tech start-ups emergence, to enhance understanding of how to survive gestation challenges. Secondly, Cantù et al. (2018) assert the need for deeper depth of knowledge, to help mitigate current mortality rate. We discover from literature that many studies have been conducted, yet, most prior studies focus on fragmented studies of the determinants of tech start-up emergence (Shepherd et al., 2020). This suggests that most studies view tech start-up gestation as a unidimensional event (Brahma et al., 2018; Gartner, 1985), which hinders better understanding of the collective activities that are likely to be highly interdependent and contradictions that birthed tech start-up emergence. (Lichtenstein et al., 2007).

Razmdoost et al. (2020) emphasize that fragmented studies are not sufficient for an in-depth understanding of tech start-up formation. Hence, we investigate the diverse dimensions of events that reveals the aggregation of the conducted interdependent activities describing how viable tech start-up emerge. The multi-dimensional events are sequence of different happenings that emerge as founding team conduct interactive and recursive activities. This happenstance, hierarchically emerge in phases, and culminated to a viable tech start-up. Thus, to identify the series of events, and understand how the recursive activities underlying the events occur, we draw from extant literature and activity theory (AT). We adapt the six elements of the activity system, to analyze the sequence of unpredictable dimensions of events as we ask the following research question (RQ): *What are the multi-dimensional events that describe viable tech start-up emergence in Nigeria tech start-up ecosystem?* The outcome of this study contributes to literature, advances theory, enlightening nascent entrepreneurs on the activities and events that are critical, and further enlighten policy makers' in developing right policies.

The remaining parts of this paper is structured as follows: firstly, we review literature for theory background of the investigation. Secondly, we select and describe the methodology appropriate for the investigation and we analyze the collected data set for the study. Thirdly, we present our findings from the dataset. Fourthly, we engage in discussion of the findings, finally, we conclude by describing contributions of the study to existing body of knowledge, in practice and in advancing theory, and further render suggestions for future study.

## 2. THEORETICAL BACKGROUND

We adopted activity theory (AT), because it possesses the power to describe the "complexity [experienced] in human activities" (Hite & Thompson, 2019, p. 2 ). Activity theory originated from a Russian scholar and psychologist called Lev Vygotsky, whose model is the first generation activity theory. However, many generations have emerged, and we draw from Yrjö Engeström perspective, called the third generation of activity theory (Engestrom et al., 1999; Madyarov & Taef, 2012). This is because it enables interactivity between multiple activity systems that culminates to a common outcome. Focusing on our phenomenon, it enables the description of the recursive activities transformation to multi-dimensional events represented as object of activity.

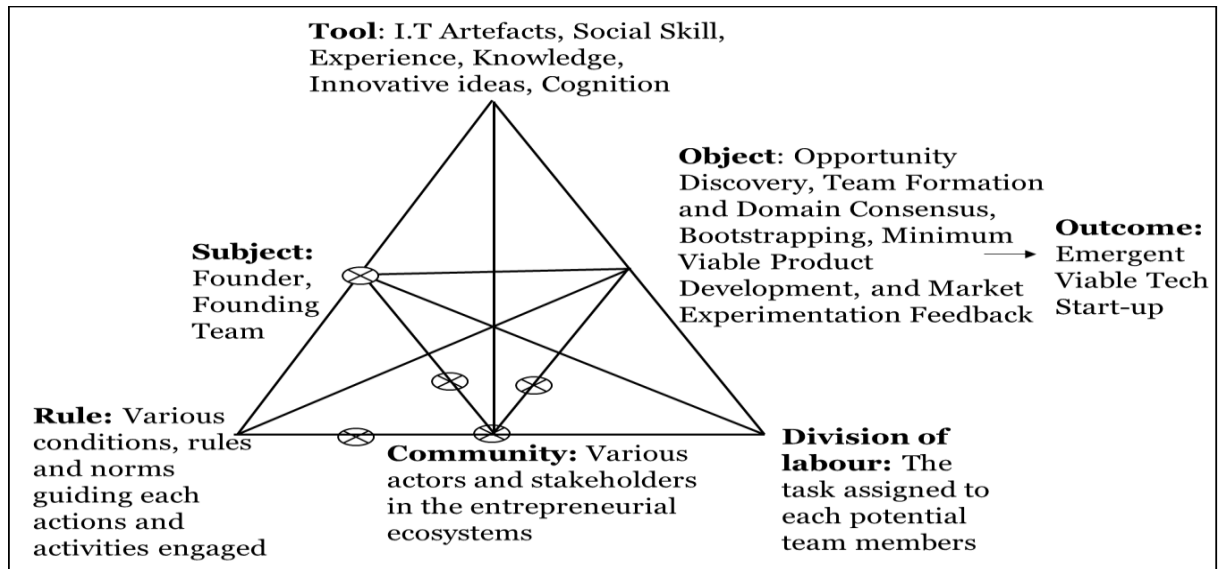

**Figure 1: Activity System Describing Tech Start-up Emergence, as adapted from (Engestrom et al., 1999)**

Thus, we use AT to describe various interactive and recursive gestation activities causing each dimension of event, by mapping founding team activities to corresponding interactive elements in the activity system as shown in Figure 1. Activity systems are guided by some fundamental principles (Bharosa et al., 2012; Engestrom et al., 1999), which include: (1) activity system, identified as the unit of analysis (Bharosa et al., 2012; Murphy & Rodriguez-Manzanares, 2008). (2) the activity system is exposed to diverse perspective and views, leading to multiple voices and choice of interest in the activity systems, and is caused by division of labor, thus, creating multiple view from the actors (Madyarov & Taef, 2012; Murphy & Rodriguez-Manzanares, 2008). (3) activity system is dependent on their own history, as they engage in the analysis of human activities through mutual influence between human actions and institutional structures (Bharosa et al., 2012; Madyarov & Taef, 2012). (4) contradiction plays a critical role as it generates changes and tensions within the activity systems to drive desired outcome (Bharosa et al., 2012; Madyarov & Taef, 2012). (5) activity systems have the potential for expansive possibility (Engestrom et al., 1999; Murphy & Rodriguez-Manzanares, 2008). The interactive activities triggers contradictions manifest as disturbances, tensions and conflicts, within or between activity systems. Such contradictions confronts nascent entrepreneurs within and between activity systems, influencing founders actions that leads to desired transformations in a system (Madyarov & Taef, 2012). Examples of contradictions in activity systems include primary, secondary, tertiary, and quaternary levels of contradictions (Madyarov & Taef, 2012). Where "the primary contradictions arise within the elements/nodes of the activity system … Secondary contradictions occur between the nodes of an activity system, and tertiary and quaternary contradictions occur between different activity systems" (Madyarov & Taef, 2012, p. 81).

Particularizing AT to our phenomenon of interest, we adapt the activity system to describe how founding team interacts with actors in the ecosystem; how they orchestrate activities that drives the sequence of multi-dimensional events as object of activity that culminates to tech start-up. Literature has shown that tech start-up emergence is a progressive gestation process, that is characterized by unpredictable multi-dimensional events (Gartner, 1985; Reynolds & Miller, 1992). Such events include opportunity emergence (Jones & Barnir, 2018), behavior and cognitive factors (Baron, 2007), opportunity confidence (Dimov, 2010), founding team experience, and industrial context (Li & Dutta, 2018), business planning in venture emergence (Long et al., 2016), and action place in venture creation (Gordon, 2012). This amplifies the need for aggregated in-depth investigation of the culminating multi-dimensional events involved. We illustrate our findings from literature in Table 1, depicting the outcome of our systematic literature review. Thus, we conceptualizes tech start-up emergence into four phase events. Such events include opportunity discovery and selection, team formation and domain consensus, bootstrapping, and MVP development and market experimentation feedback.

| **Culminating Events as Object of Activity System** | **Sources** |
|---|---|
| **Opportunity Discovery and Selection**<br>-Economically viable opportunity is the reason entrepreneurs indulge in entrepreneurship (Politis & Gabrielsson, 2006).<br>-Opportunity discovery is the prerequisite for entrepreneurship (Park, 2005).<br>-"Discovery cannot occur without [adequate] information" (Fiet & Patel, 2006, p. 216 ). Therefore, information triggers opportunity discovery, which leads to the creation of a tech start-up (Ford & Sullivan, 2008). | (Atherton, 2007; Baron, 2007; Becker & Dodo zu, 2015; Carolis & Saparito, 2006; Dimov, 2010; Fiet & Patel, 2006; Gruber et al., 2008; Harper, 2005b; Jones & Holt, 2008; Jones & Barnir, 2018; Park, 2005; Politis & Gabrielsson, 2006) |
| **Team Formation and Domain Consensus**<br>-To achieve emergence of a viable tech stat-up in the face of environmental challenges, it requires the effort of an entrepreneurial team, rather than the effort of a single entrepreneur (Diakanastasi & Karagiannaki, 2016). | (Baron, 2007; Becker & Dodo zu, 2015; Diakanastasi & Karagiannaki, 2016; Ford & Sullivan, 2008; Middleton & Nowell, 2018; Samalopanan & Balasubramaniam, 2020). |
| **Bootstrapping**<br>-A sequence of actions taken by a founding team, to enable them access necessary resources, capable of influencing the transformation of discovered opportunity to a valued product/services in a target market (Patel et al., 2011). | (Patel et al., 2011; Perry et al., 2011; Waleczek et al., 2018). |
| **MVP Development and Market Experimentation feedback**<br>-Minimum viable product (MVP) emerges from the sets of product development activities carried out by the founding team (Carmine et al., 2014). Developed MVP is tested in the market to ascertain its acceptance, reasonable traction, and further obtain feedback for possible product modification, with the aim of meeting the needs of the target customers (Ganesaraman, 2018). | (Adamczyk, 2017; Becker & Dodo zu, 2015; Bhave, 1994; Carmine et al., 2014; Ford & Sullivan, 2008; Isabelle et al., 2016; Morris & Kuratko, 2020). |

**Table 1: Tech start-up Emergence as an Outcome of Culminating Events**.

## 3.  METHODOLOGY

"Research conducted within the activity theoretical framework calls for a research design that would afford an in-depth study of an activity system or a constellation of interacting activity systems" (Madyarov & Taef, 2012, p. 84 ). This informs our choice in selecting case study as the appropriate methodology for this investigation (Walsham, 1995; Yin, 2018 ). Thus, it allows us to investigate series of events that are context-based and real life situated (Benbasat et al., 1987).

### 3.1. Case Description

Nigeria tech start-up ecosystem provides platform for socio-economic development of Nigeria. It hosts technology hub that provided support to founders and tech start-ups. Examples of the technology hubs include CC-hub, Hub-one, Next-hub, and Venier-hub. The technology hubs provide some services to the founders during tech start-up gestation; such services include accelerator services, incubator services, and workspaces to aid infrastructural needs of the founders. These hubs are mainly clustered in Lagos, the commercial capital of Nigeria. Making Lagos a tech-based entrepreneurial ecosystem and attractive city to nascent entrepreneurs and investors. It is worth noting that the large addressable market in Nigeria is responsible in attracting interest both locally and internationally ecosystem. However, the ecosystem experiences environmental uncertainty and resource constrain, which explains why only few states out of many states in Nigeria are hosting tech start-ups in their numbers within Nigeria tech start-up ecosystem. Yet, the number of tech start-ups that succeeded are few as record shows high failure experiences, at the rate of 61% (Onoja, 2020). The high rate of failure is due to poor infrastructure, scarcity of funds, political instability, high Government tax, unfavorable policies, high barrier to entry and lack of adequate skilled and experienced professionals. However, It is on record that Nigeria tech start-up ecosystem has experienced tremendous growth and recognition in Fintech, Healthtech, and Agritech, across the globe in the last five years (StartupUniversal, 2020).

### 3.2 Data Collection

The data used in this study emanated from three tech hubs (i.e. CC-Hub, Next-Hub, and Hub-one), which was selected within Nigeria tech start-up ecosystem. We conducted a semi-structured interview as primary source of our data, with the respondents consisting of tech start-up founders and co-founders. We further engaged in observation of the activities carried out by the founding team, and accessed the corresponding archival documents available, to enhance the data collected.

| S/N | Tech Start-up | IT Artifact Developed as Product/Service | Interviewee Position | Number of Years On | Technology Hub |
|---|---|---|---|---|---|
| 1 | AA | Retail Lender (Fintech) Application | Founder | Three (3) | CC-Hub |
| 2 | BB | Shopping Mall Advertising Appl. | Founder | Five (5) | CC-Hub |
| 3 | CC | Digital Marketing Application | Co-founder | Four (4) | CC-Hub |
| 4 | DD | Software Consulting Application | Founder | Four (4) | CC-Hub |
| 5 | EE | Art Market Application | Founder | Three (3) | CC-Hub |
| 6 | FF | Ecommerce Application | Co-founder | Three (3) | Next-Hub |
| 7 | GG | Financial Service Application | Co-founder | Three(3) | Next-Hub |
| 8 | HH | Property- Tech Application | Founder | Two (2) | Hub-One |
| 9 | II | Ecommerce Solution | Founder | Three (3) | CC-Hub |
| 10 | JJ | Digital Media Application | Co-founder | Four (4) | Hub-One |
| 11 | KK | Fintech (Osusu) Application | Founder | Four (4) | Hub-One |
| 12 | LL | Financial Advisory Application | Co-founder | Three (3) | Hub-One |

**Table 2: Sample Respondents of tech start-ups interviewed in the empirical situation.**

During participant observation, we observed participants in three selected tech hubs consecutively, for a period of 6 months in two phases of 3 months each and we recorded our observation in our field notes. The process of data collection lasted for nine (9) months, from January to September 2020.

We conducted interviews for twelve (12) people consisting of founders and co-founders. Each of the interview took an average of forty-five (45) munities. The questions asked during the interview were open ended, which encouraged the interviewees to express in detail their view regarding series of actions taken as they engage in recursive activities (Walsham, 1995). Some of the questions we asked include:

*What motivated you to engage in creating a tech start-up? What are the key activities for the entire process? How did you develop the product/services for the tech start-up? How did you acquire the necessary resources needed for the development of tech start-up?*

The interview proceeded with interviewees introducing and elaborating their tech start-up goals, how the idea were discovered, the development process, team formation, fund raising, market experimentation, and the kind of value their products provided to the expected customers. Table 1 presented the details of the tech start-ups that were investigated. However, the names of the tech start-ups where replaced with letters for anonymous purposes and to strengthen confidentiality.

## 4. DATA ANALYSIS AND FINDING

We focused on the collected data, transcribed the audio data to English text, and engaged in analyzing the transcribed audio data using thematic analysis (Braun & Clarke, 2006). Themes, and overarching theme where identified and properly grouped. Adopting AT, enables the investigators to engage activity system in providing description of the emergent sequence of critical events (Hardman, 2005). Our findings indicate that the events that culminate to viable tech start-up, emerges as a transformation of the action taken by founders towards resolving various tensions and disturbances arising from contradictions experienced in the activity systems. The experienced contradictions manifested among founders, founders and stakeholders, and from market experimentation feedback. Table 3 below depict contradictions experienced at each phase, and how the founding team orchestrate actions to resolve them.

| Contradictions experienced in the activity systems | Excerpts from interview | Emergent Events in Phases | Outcome of Culminated Events |
|---|---|---|---|
| Emanate from divergence views of founders in activities conducted when evaluating discovered opportunities, This generates tensions that led to the team orchestrating actions carried out to resolve tension, thus, leading to the emergence of economic valued opportunity. | *"For us in EE to arrive at our business idea, it took a lot from each of the founding team, because we had some disagreement on which product will have high traction, we then engage in further market research to reach a consensus".* | Phase:1 Opportunity Discovery and Selection Event | |
| Tensions are being experienced among the team members while conducting activities, where a team member is influenced by | *"I had to let a team member live the team at the beginning of this start-up because he fill,* | Phase:2 Team Formation and Domain | A Viable Tech Start-up Emergence |
| the magnitude of task they engage, and in turn, seek for review of equity offer allocated to them. The team resolves it by replacing such member with a more focused and goal fit individual, thus, the emergence of right team. | *he is offering so much for the percentage of equity I offered him, he started creating problem, and that was not healthy for us, so I had to let him go".* | Consensus Event | |
| This event emerge as founders take actions to resolve tension arising from limited resources, and disturbance from family and friends who decides to change their mind and request for the fund they invested. | *"My cousin created issues, when he demanded for profit or we return his money after one year of investment, the team acted by seeking for new angel investor".* | Phase:3 Bootstrapping Event | |
| Disturbance arises during product development, caused by task conflict between the founders when executing task. Another tension arises from market feedback, which drives the desired changes and modification of the product to meet the needs of the customers. Thus, the founders orchestrates action to resolve this tension, which leads to the emergence of a viable product in the market. | *"The first time we introduced our prototype to the market, the comments almost discouraged us, but we hold on to the fact that some people said they need the product, if we can modify some aspect".* | Phase:4 Minimum Viable Product Development, Market Experimentation Feedback | |

**Table 3: Contradictions as Driving Force for Tech Start-up Emergence**

Founders' engages in experiential learning from the feedback of conducted market research and prototype experimentation. Experiential learning provides critical information that act as driving force for continuous iterative activities, which play a critical role in resolving generated tensions and disturbances experienced. Thus, focusing on Figure 2,3,4,5 we describe in detail the activities driving the multi-dimensional events that culminate to viable tech start-up emergence by starting with opportunity discovery and selection of events.

### 4.1    Phase 1: Opportunity Discovery and Selection Events

**Subject (Founding Team):** The founding team is responsible for the series of iterative actions taken to search and evaluate opportunities that are transformable to innovative products/services. The founders focus on the prevailing challenges in Nigeria market to discover opportunities, which they evaluate for economic value. Excerpt, *"We began by searching, we identified a problem, and we desired to solve it. As we keep building solutions around it, we try to be very close to our customers, we try to understand how they like to be served, what exactly they need, so we can make necessary adjustment to keep re-inventing our product to fit the market".*

**Rules:** The founders ensure they keep necessary government regulations and policies guiding the sector of interest. In some cases, conditions placed on some business makes it difficult for them to venture. Thus, founders collaborate with people who can provide those requirements. Excerpt,

*"You also need to look at the regulatory constrains for the discovered business idea, and how to meet those regulatory requirements".*

*"Regulatory requirement can sometimes be an impediment ... We partnered with people with the financial resources; in return we offer them percentage of the venture equity".*

**Community:** This includes founders, customers, business advisers, mentors, investors, government institutes, and competitors. The founders approach potential customers in the target market, to discuss and scrutinize opportunities discovered, by interacting with them to understand existing challenges, and identify how to transform the opportunities discovered. Furthermore, the founders interact with stakeholders within the ecosystem, for further evaluation. Excerpt, *"We are creating a world where everyone and anyone can invest in real estate through block chain technology. We are working with several real estate partners and stakeholder in the sector"*.

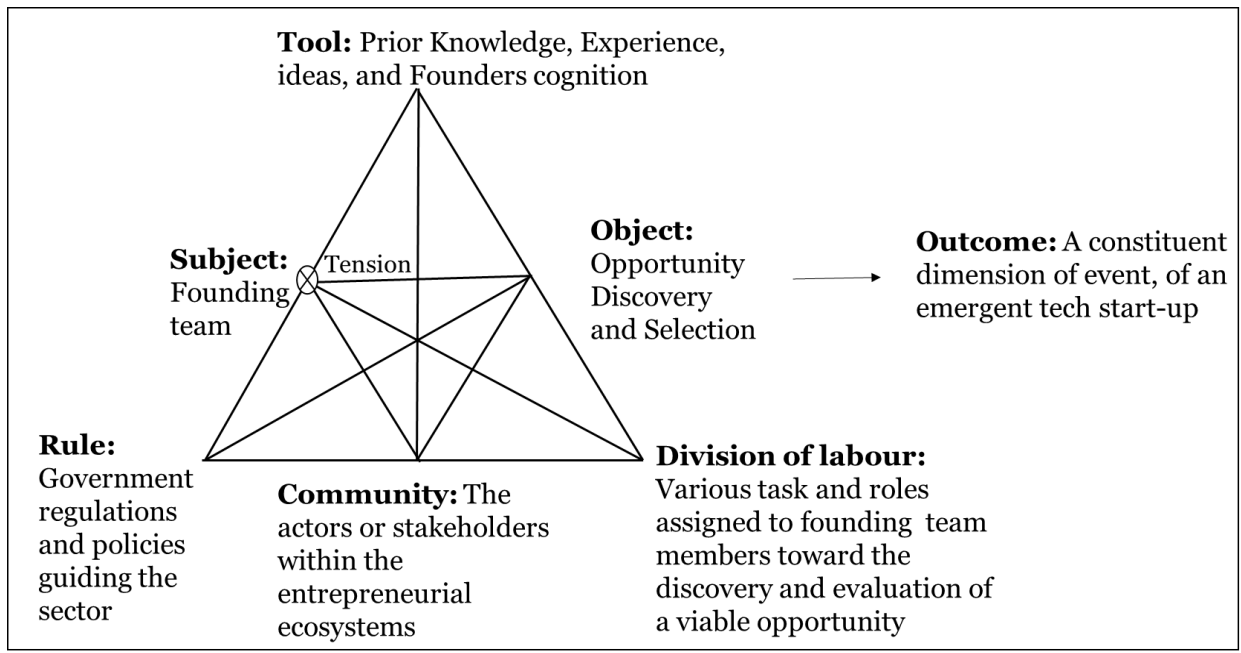

**Figure 2: Activity System for Opportunity Discovery and Selection**

**Tool:** The founders uses cognitive abilities, prior knowledge and experiences, as tools within the specific domain of interest, as they search for marketable opportunity. Excerpt, *"We all have domain experience in what we do; it was easy for us to navigate all the problems we probably have faced as we search for marketable opportunity within our environment. So experience also is key"*

**Division of labor:** The founder engages services of experts, business advisors, and skilled workers to evaluate the identified opportunities. Such that some are responsible in making enquiry from the prospective customers, on how tenable a particular opportunity is to the target market. Another is responsible in converting this idea to an innovative product design, while some other are responsible for developing business plan and marketing strategies needed. Excerpt, *"There is a lot to benefit from synergizing and exchanging expertise. So I think in this tough environment, attempting to do it alone might not be enough"*.

## 4.2    Phase 2: Team Formation and Domain Consensus Event

**Subject (Founding Team):** In this phase, the founding team begins to search for complementary team members needed to transform the selected opportunity to a product fit for the target market. Illustrating, *"It is important for the founder of a tech start-up to form a team, by engaging in partnerships that saves you money and time. Engage in partnership that opens door for you, links you up with other bigger organization"*.

**Rule:** There are conditions for partnership, the founders engage in reaching agreement with the prospective member on the equity offer, agree on terms and condition of work, and agree on contribution they are bringing to the team. Then, they seal the alliance with a sweat equity, a percentage of equity for collaborating. Excerpts, *"Number one thing for me is to have a very good* *team and ensure fairness in equity allocation to members of the team. I do not hoard equity from the team. For example, you find some founders who want to hold 80% of the equity and they are been very stingy with what they can give out and that in turn affect what they can achieve".*

**Community:** The team consist of various personnel who compliments each other in expertise and in finance. It involves business advisors, highly skilled people, experience people, and expatriates in the domain of interest. Excerpt, *"I feel like I did not do it alone, I had a partner at a time and when we collected money from people, they became partners too. Therefore, it made it more of a teamwork and not just an individual thing".*

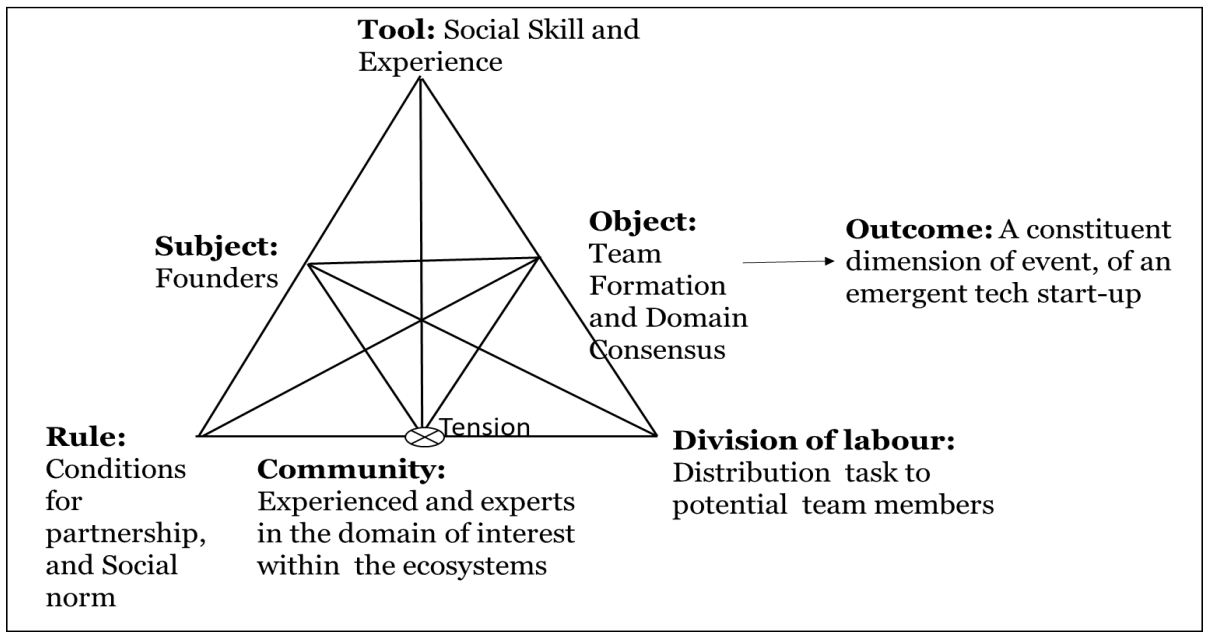

**Figure 3: Activity System for Team Selection and Consensus.**

**Tool:** The founders uses their social skills and experiences as tools in convincing prospective team members to join the team. Excerpt, *"I had to bring in a tech expert to become a co-founder, which enables me to understand the technical side over time to grow my skills".*

**Division of labor:** For instance, *"As an entrepreneur, you need to surround yourself with a solid network of associates, people who are more experienced than you in different field. I am lucky to have such kind of support structure too".*

### 4.3   Phase 3: Bootstrapping Event

**Subject (Founding Team):** The founding team as subject engages in bootstrapping activities to acquire necessary resources needed to transform the selected opportunity to a viable tech start-up. They directly invest their personal funds and interact with families and friends to persuade them to invest funds, and other kinds of resources needed to achieve viable tech start-up. Excerpt, *"Am not a fan of I want to take 100% control of my business because when we started, I realized the level of my strength, that I can't do it all by myself and that I would be needing help from others. So I immediately reached out to the people am going to need their help, which cost some equity, which is not a problem for me, because I would rather own 20% of a 10-Billion-dollar business than own a 100% of 1-million-dollar business".*

**Rule:** To bootstrap successfully, the founding team focused on ensuring that cultural values and norms binding them to their families and friends are maintained, also personal commitment to ensure trust are established and strengthen. Excerpt, *"If an idea comes to your head … you have to put your own skin in the game. If you are a technical founder who build the product, that might be your own equity commitment in the business, it might not really be cash because you spent sleepless nights in building the product"*

**Community:** These includes family members, friends, skilled and technical personnel, who are willing to offer their resources to support the development of the tech start-up. *"You have to identify skilled and technical experts who believes in your idea; you also have to be in the good books with family and friends because they are usually the first point of call when you are cash trapped".*

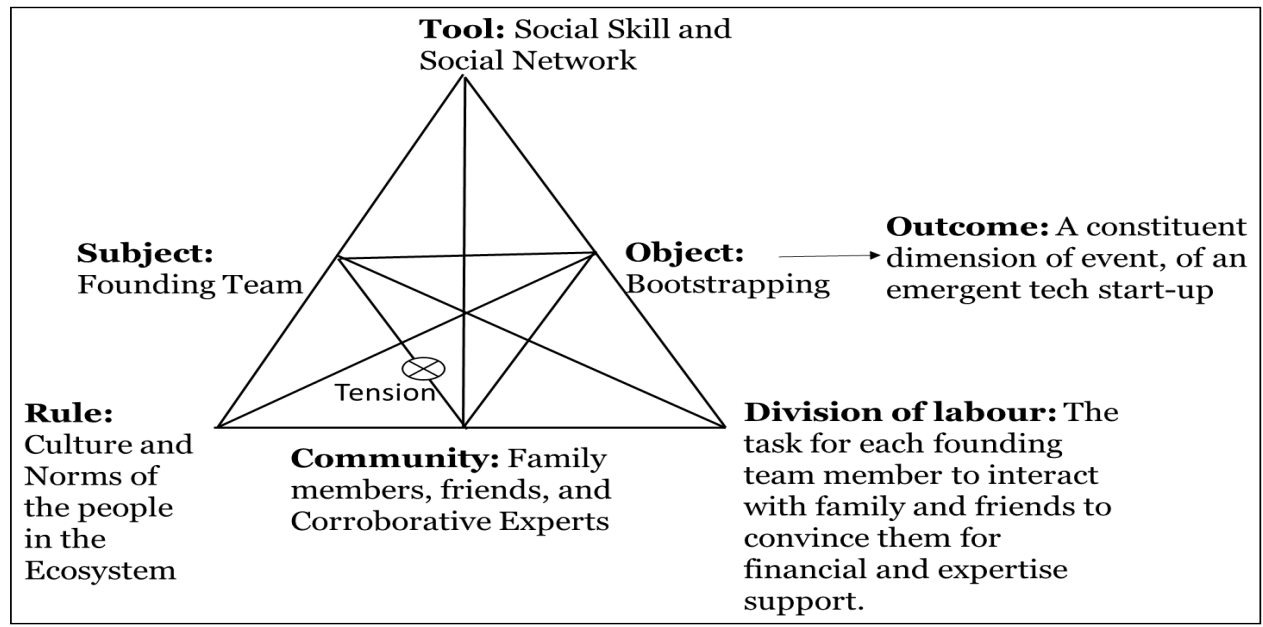

**Figure 4: Activity System for Bootstrapping.**

**Tool:** The founders uses social skills and social networks within and outside the ecosystem, by engaging in profitable interaction and association with family members, friends and colleagues. Excerpt, *"Leveraging on people is critical in this environment, if you cannot leverage on people and get things done free or for differed payment or at a subsidized rate, then you will not be able to do it as a lean start-up".*

**Division of Labor:** Each member of the founding team is tasked to convince and attract rich and skilled family members or friends that can invest finances or skills towards the development of the tech start-up. Excerpt, *"I feel like I did not do it alone, I had a partner at a time and when we collected money from people, they became partners too. Therefore, it made it more of a teamwork and not just an individual thing".*

### 4.4　　Phase 4: Minimum Viable Product Development, Market Experimentation Feedback

**Subject (Founding Team):** The founders are responsible in setting up a team, technically capable of developing product/service prototype, which later undergoes acceptance and viability test in the market. Minimum viable product (MVP) development emerges from the recursive activities carried out by the team by engaging technical expertise in developing a prototype. After developing the prototype, it is experimented in the market to ensure viability, possible traction, and obtain feedback for necessary modification. Excerpt, *"The founder conceptualizes the ideas and then build a sort of Minimum Viable Product (MVP) that is tested in the target market".*

Thus, during the initial entry to market, the founding team interacts with the market forces by sampling continuously the MVP and proposed customers to understand the needs of the market (Morris & Kuratko, 2020). Consequently, market experimentation, allows the founders to engage in "learning from what does not work, and experimenting with alternative approaches, the entrepreneur is able to discover where the real opportunity lies" (Morris & Kuratko, 2020, p. 8 ). *"Founders need to be observant of the market and the product, understand your market and try to ensure that product is fit for the market, and the market wants it. By engaging on a proper market survey before developing the final product".*

**Rule:** The rule observed in this event are the ideal practices of professional code of conduct (i.e. formal practices) in developing IT artifacts. Another rule is ensuring constant communication with target customers during IT artifact development. The next rule observed is the government policies guiding the development of a specific product, especially in the area of quality control and quality assurance. For instance, developing financial service Application (e.g. Osusu App.), requires founders ensuring that policies guiding financial services are adhered to as the IT artifact is been developed. Excerpt, *"When starting to develop a product for a business, you must not stop the product development because of unfavorable policy guiding the business. However, if it is a highly regulated sector, please ensure to get the license needed for you to operate in that sector"*.

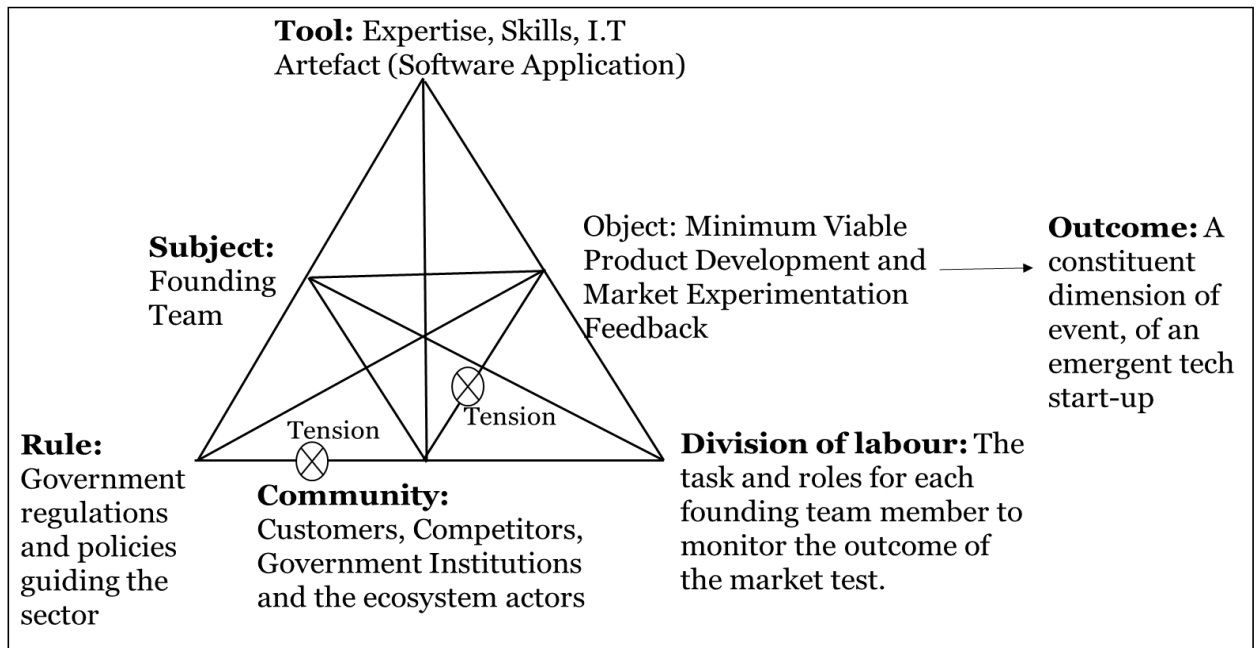

**Figure 5: Activity System for MVP Development and Market Experimentation Feedback**

**Community:** The community include founding team who develops the product, prospective customers who test and provide feedback to the founders on how the MVP meets their needs with necessary modification. Again, part of the community are government institutions regulating the specific sector the founders are venturing, and other stakeholders within the ecosystems. Excerpt, *"It is a whole lot of energy that has to go into market validation, R&D to ensure that the solution you are providing is much needed, and not just assuming that it is needed. You must be able to proffer solution to local challenges, which meets government regulations, which can stand competitors product, and can easily attract investors in the immediate community"*.

**Tool:** The tools are expertise, skills, and developed IT artifact introduced to the market. Excerpt, *"Another thing is as much as possible; you continue to develop and modify your MVP till you have customers. Nothing speaks for itself like having customers who are using your product or service"*.

**Division of Labor:** Tasks are being distributed to the team members. For instance, some are responsible in coding, fixing of any observed bug, and modification of code for new features; some members of the team are responsible in physically managing the development process of the product; some are responsible in communicating with the customers, as the MVP is been tested in the market. While some are responsible in analyzing feedback from the market; and others are responsible in ensuring that the outcomes of the feedback are properly implemented to suit the desires and needs of customers. Excerpt, *"Developing a start-up is more than having a great idea; it is being able to come up with the right team that can actually execute the business idea in to a viable product ... Hence, you need to get a group of people with requisite knowledge, experience and expertise that can actually translate your idea to something tangible and acceptable to the target market"*.

# 5. DISCUSSION

The outcome of our empirical investigation suggest that tech start-up gestation consist of four critical phases, catalyzed by continuous feedback from the market. Our findings reveal how tech start-ups emerge from a hierarchical sequence of events, driven by interdependent recursive activities conducted by the founding team and other stakeholders in the ecosystem.

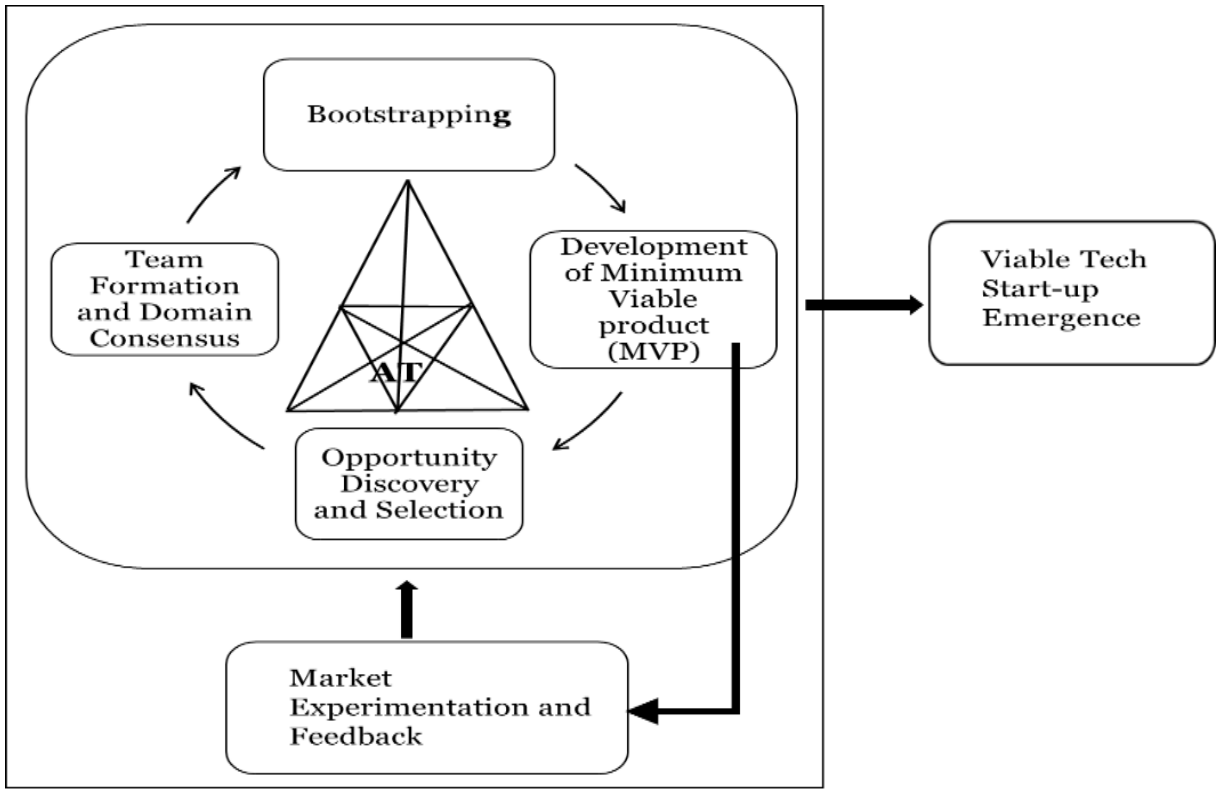

**Figure 6: Multi-dimensional events culminating to emergent Viable Tech Start-up.**

Our findings in phase 1 suggest that opportunity discovered in a target market is a prerequisite for entrepreneurship (Park, 2005), and does not support founders creating innovative opportunity from their thought and prior experience. Therefore founders in our context, embrace only discovered opportunity as raw material required to achieve entrepreneurship (Baron, 2007), and the reason they indulge in entrepreneurial activities (Politis & Gabrielsson, 2006). They evaluate and scrutinize the identified opportunities to ensure they select opportunity with true economic value (Harper, 2005a). This outcome corroborated the result of some prior fragmented studies that focus on opportunity recognition (Bhave, 1994). However, this result is only a fragment of the dimensions of events that constitute tech start-up emergence. The second phase involves bringing together a group of people that possess complimenting skills, with the aim of collectively applying domain knowledge to creating a viable tech start-up (Middleton & Nowell, 2018). Our finding shows that appropriate team formation is another dimension of event that is critical to the emergence of a viable tech start-up (Diakanastasi & Karagiannaki, 2016). As we discovered in our context, it is practically difficult for a single founder to develop a tech start-up because the context is plagued with uncertainties caused by various market forces. More people are usually involved, to enable the provision of needed knowledge, other resources and necessary relationship that will facilitate meeting various environmental demands necessary for the emergence of a viable tech start-up (Ford & Sullivan, 2008). Emphasizing further, our findings support Diakanastasi and Karagiannaki (2016) who argue that achieving emergence of a viable tech stat-up requires the effort of entrepreneurial team, rather than the effort of a single entrepreneur. Therefore, each member of the team provides skills, experiences and capabilities that compliments the short fall of others within the team (Middleton & Nowell, 2018).

Thirdly, resource acquisition plays a key role in tech start-up emergence. Its absence have led to failure of many tech start-ups during gestation (Baron, 2007). Founders often face challenges for necessary resources in our context. Thus, phase three reveal the key process of gaining access to needed resources (i.e. financial and humans) without engaging in borrowing that requires collateral in our context (Patel et al., 2011). Thus, founders engagement in "bootstrapping activities used either for venture development or product development" (Patel et al., 2011, p. 423 ), and it includes the acquisition of needed resources from "relatives and friends, using credit cards, withholding salary, working in other businesses, and employing relatives for nonmarket-based salary" (Waleczek et al., 2018, p. 3 ). In the final phase, our result shows that minimum viable product (i.e. Prototype) development is the outcome of activities carried out by the founding teams with the aim to provide product/service that meets the unmet needs of the market (Carmine et al., 2014). The developed MVP is experimented in the market, to create learning opportunity for the founders, to ascertain the product/service viability in the market. Our finding support the view that market experimentation activity is critical in establishing communication between founders and the customers concerning the product under development (Adamczyk, 2017; Ganesaraman, 2018). In addition, it helps to solve the issues resulting from customers' choice dynamics. Consequently, received feedback from the market enables founders to ensures that "adjustments are being made as the individual figures out what works in practice and comes to better understand the market opportunity" (Morris & Kuratko, 2020, p. 3 ). Therefore, initial entry to market enables the founding team to interacts with market forces by sampling continuously the prototype, and proposed customers to understand their unmet needs (Morris & Kuratko, 2020). However, our finding shows that acceptance of a product can only be, if it is being discovered from the existential challenge in the market. This suggest that creating product without properly engaging initial market research will lead to tech start-up failure in our context. As the events interact, they create contradictions that lead to the transformation of the desired outcomes (Madyarov & Taef, 2012).

## 6. CONCLUSION AND FUTURE STUDIES

The outcome of this study contributes to theory, and in practice. It provides description of tech start-up emergence in Nigeria. In practice, this study is relevant, as it enlightens nascent entrepreneurs on how to act during development of a viable tech start-up in an uncertain market. Further, it added empirically, to existing literature, by enhancing the narrative regarding fragmented studies on the determinants of tech start-up emergence. By exploring cumulatively, the sequence of multi-dimensional events characterizing the emergence of a viable tech startup in an uncertain environment. Future study will seek to identify causal mechanisms that must exist for the events to occur.